\documentclass[prd,showpacs,eqsecnum,twocolumn]{revtex4}
\usepackage{bm}
\usepackage{amssymb}
\usepackage{amsmath}

\begin{document}
\title{Multipole structure of current vectors in curved spacetime}
\author{Abraham I. Harte}
\affiliation{Institute for Gravitational Physics and Geometry and Center for Gravitational Wave Physics,
\\
The Pennsylvania State University, University Park, PA 16802, USA}

\date{November 13, 2006}

\begin{abstract}
A method is presented which allows the exact construction of
conserved (i.e. divergence-free) current vectors from appropriate
sets of multipole moments. Physically, such objects may be taken to
represent the flux of particles or electric charge inside some
classical extended body. Several applications are discussed. In
particular, it is shown how to easily write down the class of all
smooth and spatially-bounded currents with a given total charge.
This implicitly provides restrictions on the moments arising from
the smoothness of physically reasonable vector fields. We also show
that requiring all of the moments to be constant in an appropriate
sense is often impossible; likely limiting the applicability of the
Ehlers-Rudolph-Dixon notion of quasirigid motion. A simple condition is also derived that allows currents to exist in two different
spacetimes with identical sets of multipole moments (in a natural
sense).
\end{abstract}


\maketitle

\vskip 2pc

\section{Introduction}

There are a number of instances of ``currents'' appearing throughout
physics as fields which somehow express a conservation law. Here, we
will be studying vector fields $\mathbf{J}(x)$ in $3+1$ dimensional
spacetimes. In classical physics, a common example of such an object
would be the electromagnetic current. Another would be the numerical
flux density of some species in a continuum (as long as the total
number of such particles isn't expected to change via chemical or
other reactions).

Keeping these examples in mind, we take the support of such currents
to be a spatially-bounded region $W$ which surrounds a timelike
worldline $Z$. Intuitively, this could be thought of as the
worldtube of some extended body. The ``central'' worldline would
then be most naturally identified with the body's center-of-mass.

We may also use any spacelike hypersurface $S$ which cuts $W$
to define a charge in the usual way:
\begin{equation}
q(S) := \int_{S} \mathbf{J} \cdot \mathrm{d}\mathbf{S} ~.
\label{chargeDefine}
\end{equation}
The current would be ``conserved'' if $q(S)$ were independent of $S$.
This clearly occurs when
\begin{equation}
\nabla \cdot \mathbf{J} =0 ~. \label{DivFree}
\end{equation}

Our purpose in this paper is to study vector fields which satisfy
this equation and have the given support properties. In a sense,
finding such fields is actually quite simple. This can be seen by
constructing a $3$-form from $\mathbf{J}$ in the natural way, and
then observing that (\ref{DivFree}) is equivalent to requiring that
this form be closed. Solutions follow immediately. For example,
\cite{Elast} discusses a common procedure in continuum mechanics
whereby the number density flux is essentially the pullback to
spacetime of a fixed volume form defined in a three-dimensional
``body space.'' This method is quite natural when discussing the
worldlines of a body's constituent particles, but may be rather
awkward in other cases.

For example, it is often useful to decompose a current into a set of
multipole moments. In some cases, moments may be experimentally
determined be measuring the asymptotic behavior of the fields
sourced a body. If $\mathbf{J}$ happened to be an electric current,
the moments would also appear as coefficients in an expansion for
the net force and torque due to externally applied electromagnetic
fields \cite{Harte, Dix67, Dix74}. Even when including the effects
of self-fields, some aspects of the multipole decomposition remain
useful in a wide range of scenarios \cite{Harte, PoissonExtend}.

These types of methods also provide an elegant way of separating out
different scales when discussing the structure of some extended
body. Since the moments (that we will choose) are in fact tensor
fields defined along $Z$, it becomes relatively simple to discuss
this structure from the point of view of a single comoving observer.
This makes a number of problems more intuitive.

Unfortunately, multipole moments are related to $\mathbf{J}$ (or the
deformation gradient which may be used to construct it) in a rather
complicated way. In this paper, we shall take the point of view that
the moments themselves are fundamental. Other more
commonly-discussed quantities are to be derived from them. It is
shown how to explicitly construct $\mathbf{J}$ in this way in Sec.
\ref{ConstructCurrent}, which leads to the introduction of two
``moment potentials'' defined on the tangent bundle restricted to
$Z$. A moment may be derived from these quantities essentially by
taking a (three-dimensional) Fourier transform, and then
differentiating a sufficient number of times. Importantly, these
potentials are not required to obey any evolution equations. Their
``time-dependence'' is essentially arbitrary.

Not all of them will describe physically reasonable currents,
however. Most choices introduce singularities. Sec.
\ref{ChooseCurrent} discusses how to avoid such problems. This
implicitly shows which restrictions must be placed on the moments
for them to represent plausible current vectors.

As a byproduct of this construction, we obtain a simple method of
constructing the class of essentially all smooth currents with a
given total charge. The situation is nearly identical to the one in
flat spacetime, where the charge density and three-current observed
by someone on $Z$ are very simply related to the moment potentials.
This could be useful in studying corrections to the Dewitt-Brehme
expression \cite{PoissonRev} for the electromagnetic self-force on a
charged test particle in curved spacetime (which is intended to
describe all ``sufficiently small'' charges with a given $q$). The
flat spacetime analog of this idea was carried out in \cite{Harte}.

Lastly, we investigate some consequences of imposing simple
\textit{a priori} conditions on the moment set itself. Sec.
\ref{Rigidity} supposes that the moments are all constant with
respect to some frame attached to the central worldline $Z$. This
closely mirrors the idea of quasirigid motion advanced in
\cite{Dix70a, EhlRud, Dix79}. Unfortunately, it is shown that the
moments can only remain constant under certain conditions involving
the Riemann tensor and the acceleration of $Z$ (roughly speaking).

Sec. \ref{Comparisons} then looks at the possibility of two currents
-- possibly in different spacetimes -- having identical sets of
multipole moments as viewed in their respective frames. As in the
case of setting all of the moments constant, this is not always
possible. A restriction on the Riemann tensors on both central
worldlines is derived which allows these types of ``comparable''
currents to exist.

The system of multipole moments adopted here is the one developed by W.~G. Dixon \cite{Dix67, Dix70a, Dix70b, Dix74, Dix79}. The relevant portions of the formalism are reviewed in appendix \ref{DixReview}. It has a number of desireable properties without which many of the results in this paper would likely be impossible to reproduce. It also has a very elegant place in the (exact) theory of relativistic mechanics. And perhaps most importantly, the formalism also works for symmetric rank-$2$ tensors $\mathbf{T}$ satisfying $\nabla \cdot \mathbf{T} =0$ (or more generally $\nabla \cdot \mathbf{T} = \mathbf{F} \cdot \mathbf{J}$ with $\mathbf{F}$ a closed 2-form; e.g. an electromagnetic field). This paper may therefore be taken as a model for attempting to construct all possible stress-energy tensors with a given center-of-mass and net linear and angular momentum vectors.

The results here also make some use of bitensors. Appendix \ref{Bitensors} reviews the subject, and states some results used in text.

For the remainder of this paper, indices shall always be included on tensor or bitensor quantities. Those which refer to points on $Z$ will be denoted by $a$, $b$, \ldots. Indices associated with all other points in $W$ will be written as $a'$, $b'$, \ldots. Tetrad labels running from $0$ to $3$ are to be expressed in capital Roman letters from the beginning of the alphabet, while triad labels running from $1$ to $3$ will be denoted by $I$, $J$, \ldots. We choose the signature of the metric to be $-+++$.

\section{Constructing currents}
\label{ConstructCurrent}

At this point, we assume that the reader has at least skimmed the
appendices to gain familiarity with the notation and basic elements
of Dixon's formalism. These outline how conserved vector fields in
spacetime may be represented by the current skeleton $\hat{J}^{a}$;
a vector-valued distribution defined on the tangent bundle
(restricted to a neighborhood of $Z$). Dixon has actually proven
that \textit{all} well-behaved currents may be uniquely represented
in this way \cite{Dix74}.

The reverse is not true, however. One could pick moments satisfying
all of the given constraints, and yet the right-hand side of
(\ref{JtoJhat}) may not even exist. Then there wouldn't be any
current vector at all associated with that set of moments. Even when
the integrals do exist, the resulting $J^{a'}$ may be singular. Such
a result would clearly be unphysical.

The main purpose of this paper is to remove this problem. We want to
be able to specify a current skeleton in some simple way which
automatically guarantees that the associated vector field is both
physically reasonable and has a given charge $q$. We do this by constructing $J^{a'}$ explicitly in terms of $\hat{J}^{a}$, and then checking that the result is reasonable.

Our first step will be to construct current vectors from all of the
skeletons satisfying the constraints described in appendix
\ref{DixReview}. The natural starting point for this is of course
(\ref{JtoJhat}). It is convenient to convert the integral over $X$
on the right-hand of this equation into one over $x = \exp_{z} X$.
The Jacobian of this transformation is given by (\ref{Jacobian}),
from which it follows that
\begin{equation}
\left\langle \mathbf{J}, \bm{\phi} \right\rangle = \int ds \left\langle
\Delta \hat{J}^{a} H^{a'}{}_{a} , \phi_{a'} \right\rangle ~.
\end{equation}
As in (\ref{JtoJhat}), $\phi_{a'}$ is an arbitrary test function.
This equation therefore determines the current completely in terms
of its skeleton (ignoring irrelevant sets of measure zero). Assuming
that the integrals on the right-hand side are absolutely convergent,
they may be freely commuted. Taking advantage of this, one finds
that
\begin{equation}
\left\langle J^{a'} - \int ds \, \Delta \hat{J}^{a} H^{a'}{}_{a} ,
\phi_{a'} \right\rangle = 0 ~,
\end{equation}
which allows the identification
\begin{equation}
J^{a'}(x) = \int ds \, \Delta(x,z) \hat{J}^{a}(X,z)
H^{a'}{}_{a}(x,z) ~. \label{JExplDefine}
\end{equation}
Here, $X^{a} = -\sigma^{a}(x,z)$ and $z=z(s)$.

In \cite{Harte}, it was found that the constraint equations
(\ref{Constr1})-(\ref{Constr3}) imply that $\check{J}^{a}$ --
defined in the text immediately preceding (\ref{ReducedSkel}) -- has
the form
\begin{eqnarray}
\check{J}^{a}(X,z) &=& \mathcal{A}^{a}(X,z) \delta( n \cdot X) -
h^{a}{}_{b}(z) X^{b} \nonumber
\\
&& ~ \times \mathcal{B}(X,z) \delta'( n \cdot X) ~. \label{Jcheck}
\end{eqnarray}
$h^{a}{}_{b}:=\delta^{a}_{b} + n^{a} n_{b}$ is the standard
projection operator, and it is given that $n^{b} \nabla_{*b}
\mathcal{A}^{a}=n^{b}\nabla_{*b} \mathcal{B}=0$. The only other
restrictions on the moments may be written as
\begin{eqnarray}
n_{a} \mathcal{A}^{a} &=& \nabla_{*a} \left( h^{a}{}_{b} X^{b}
\mathcal{B} \right) ~, \label{A0}
\\
\nabla_{*a} \mathcal{A}^{a} &=& 0 ~.
\label{divA}
\end{eqnarray}
Since $\mathcal{A}^{a}$ does not depend on $n \cdot X$, this second
equation is really only a 3-divergence. Also note that by
definition,
\begin{equation}
\hat{J}^{a}=Q^{a} \delta(X) + \check{J}^{a} ~. \label{Jhat}
\end{equation}

Using these expressions, the $s$-integral in (\ref{JExplDefine}) can
be performed explicitly to yield
\begin{eqnarray}
J^{a'} &=& \int ds \, \delta(X) Q^{a'} + N^{-1} \Big[ \Delta
\mathcal{A}^{a} H^{a'}{}_{a} \nonumber
\\
&& ~ + v^{c} \nabla_{c} \left( N^{-1} \Delta \mathcal{B} h^{a}{}_{b}
X^{b} H^{a'}{}_{a} \right) \Big] ~, \label{Current1}
\end{eqnarray}
where the right-hand side is understood to be evaluated at the
(unique) $z$ which sets $n(z) \cdot X(x,z) =0$. Also,
\begin{equation}
N := -n_{a} \sigma^{a}{}_{b} v^{b} + \dot{n}_{a} X^{a} ~.
\label{LapseDefine}
\end{equation}
Following the notation of (\ref{vDefine}), $\dot{n}^{a} := \delta
n^{a}/ds$. In flat spacetime, this expression reduces to the lapse
of the foliation $\{ \Sigma \}$. Although it doesn't retain this
interpretation more generally, we shall still refer to it as a
lapse.

Now for $x \notin Z$, note that the integral in (\ref{Current1})
doesn't contribute to the current. Temporarily restricting ourselves
to such cases, (\ref{HDerivative}) and (\ref{DeltaDerivative}) can
be used to show that
\begin{eqnarray}
J^{a'}(x) &=& \Delta N^{-1} H^{a'}{}_{a} \Big[ \mathcal{A}^{a} +
X^{a} v^{b} \nabla_{b} \left( \mathcal{B}/N \right) \nonumber
\\
&& ~+ \left( \mathcal{B}/N \right)  \Big( n^{a} \dot{n}_{b} X^{b} -
h^{a}{}_{b} \sigma^{b}{}_{c} v^{c} \nonumber
\\
&& ~ - 2 H^{b'}{}_{b} v^{c} X^{[a} \sigma^{b]}{}_{cb'} \Big) \Big]
~. \label{Current2}
\end{eqnarray}
It is natural to split this up into components adapted to an
observer moving along $Z$. Define the charge density to be
\begin{equation}
\rho(x) :=  n_{a}(z(\tau)) \sigma^{a}{}_{a'}(x,z(\tau)) J^{a'}(x) ~.
\label{rhoDefine}
\end{equation}
Similarly, let the three-current be given by
\begin{equation}
j^{a}(x) := - h^{a}{}_{b}(z(\tau)) \sigma^{b}{}_{b'}(x,z(\tau))
J^{b'}(x) ~. \label{jDefine}
\end{equation}

Substituting into (\ref{Current2}), and using (\ref{A0}) and
(\ref{LapseDefine}), the charge density reduces to
\begin{equation}
\rho = - \Delta \nabla_{*a} \left[ h^{a}{}_{b} X^{b}
\left(\mathcal{B}/N \right) \right] ~. \label{rho1}
\end{equation}
With the exception of the overall factor of $\Delta$, this is
exactly the same form that occurs in flat spacetime. Such a
simple correspondence will not hold for the three-current, however.

Before computing it, note that $\mathcal{B}$ does not appear by
itself in (\ref{rho1}). One instead finds the ratio $\mathcal{B}/N$.
This grouping will continue to be useful when expanding the
three-current, although it is a little inconvenient as stated. We
would like this ratio to have the same arguments as $\mathcal{B}$
itself. Specifically, it should be independent of $n \cdot X$. But
$n^{a} \nabla_{*a} N(X,z) \neq 0$ even when $n \cdot X =0$. To fix
this, define a ``flattened'' lapse $\tilde{N}$ as
\begin{equation}
\tilde{N}(X,z) := N( h(z) \cdot X , z ) ~,
\label{flatLapse}
\end{equation}
and a ``normalized'' moment potential
\begin{equation}
\mathcal{C}(X,z) := \mathcal{B}(X,z)/\tilde{N}(X,z) ~.
\label{CDefine}
\end{equation}
Clearly, $n^{a} \nabla_{*a} \mathcal{C}(X,z) = 0$, as desired. The
projected divergence in (\ref{rho1}) allows $\mathcal{C}$ to be substituted for  $\mathcal{B}/N$ there, so
\begin{equation}
\rho = - \Delta \nabla_{*a} \left( h^{a}{}_{b} X^{b} \mathcal{C}
\right) ~. \label{rho2}
\end{equation}

This may be further simplified by introducing an orthonormal tetrad
$\{ n^{a}(z), e^{a}_{I}(z) \}$ defined on $Z$. Enforcing the
orthonormality conditions requires that the spatial triad evolve according to
\begin{equation}
\dot{e}^{a}_{I} := v^{b}\nabla_{b} e^{a}_{I} = n^{a} \dot{n}_{b}
e^{b}_{I} + \Omega_{I}{}^{J} e_{J}^{a} ~, \label{Tetrad}
\end{equation}
where $\Omega_{IJ}(z) = \Omega_{[IJ]}(z)$ is arbitrary, and
represents the angular velocity of the triad. It needn't have any
particular physical significance, however, so we shall leave it
unspecified for now.

If the tetrad is assumed known, we can trivially introduce a set of
(normalized) Riemann normal coordinates defined by
\begin{equation}
R^{A}(x,z) := e^{A}_{a}(z) X^{a}(x,z)
\label{RDefine}
\end{equation}
for any fixed $z$. Then an arbitrary bitensor $f(x,z)$ could be
rewritten as either (abusing the notation a bit) $f(X,z)$ or
$f(R,z)$.

This is particularly useful for objects such as $\mathcal{C}$, which
are independent of $R^{0} = - n \cdot X$. So we may write
$\mathcal{C} = \mathcal{C}(\mathbf{R},z)$, where $\mathbf{R}$
represents the spatial coordinates $\{ R^{I} \}$. Defining the
differential operator $\partial_{I}$ in the obvious way,
(\ref{rho2}) now simplifies to $\rho = - \Delta \partial_{I} \left(
R^{I} \mathcal{C} \right)$.

Recall that this expression is only correct away from $Z$.
Correcting it requires evaluating the integral in (\ref{Current1}).
This is easily done by noting that $\delta(X) = \delta(R^{0})
\delta^{3}( \mathbf{R})$. Then
\begin{equation}
\int ds \, \delta(X) Q^{a'} = \delta^{3}( \mathbf{R} ) Q^{a'} ~.
\end{equation}
Using (\ref{Monopole}), the charge density everywhere is finally
given by
\begin{equation}
\rho = \Delta \left[ q \delta^{3}(\mathbf{R}) - \partial_{I} \left(
R^{I} \mathcal{C} \right) \right] ~. \label{RhoDiverge}
\end{equation}
It is now useful to define a function $\psi(\mathbf{R},z)$ which
satisfies
\begin{equation}
\partial_{I} \left( R^{I} \psi \right) = \delta^{3}(\mathbf{R}) ~.
\label{PsiDefine}
\end{equation}
This equation will be solved in the following section. But for now, $\psi$ will be left unspecified.

We instead define
\begin{equation}
\varphi(\mathbf{R},z) := q \psi(\mathbf{R},z) -
\mathcal{C}(\mathbf{R},z) ~. \label{PhiDefine}
\end{equation}
Anywhere in a neighborhood of $W$, the charge density is now
given by
\begin{equation}
\rho = \Delta \partial_{I} \left( R^{I} \varphi \right) ~.
\label{RhoFinal}
\end{equation}

$\mathcal{B}$ (and therefore $\mathcal{C}$) was originally
defined to vanish outside of $W$. But applying Gauss' theorem to
(\ref{PsiDefine}) shows that $\psi$ cannot share this property.
$\varphi$ must therefore be equal to $q \psi$ everywhere outside of
$W$. This clearly implies that $\rho = 0$ in such regions, as
required. Beyond this restriction, $\varphi$ may be specified
arbitrarily as long as the charge density which it determines
remains nonsingular.

Implicit in these statements is the fact that any function depending
on $(\mathbf{R},z)$ may be uniquely identified with another function
depending only on $x$. In a sense, we are simply choosing to work in
a special system of coordinates. These happen to be closely related
to Fermi normal coordinates, and exactly coincide with them when
$n^{a} = v^{a}$ and $\Omega^{IJ} =0$.

In any case, we may now apply a similar analysis to the three-current. Using (\ref{Current1}), (\ref{Current2}), and (\ref{jDefine}), the tetrad components of $j^{a}$ reduce to
\begin{eqnarray}
j^{I} &=& \Delta N^{-1} \Big[ \mathcal{A}^{I} + q v^{I} \delta^{3}(
\mathbf{R} ) + R^{I} \dot{\mathcal{C}} +  \partial_{J} \left( R^{I}
\mathcal{C} \right) \nonumber
\\
&& ~ \times \left( \Omega^{J}{}_{K} R^{K} - e^{J}_{b}
\sigma^{b}{}_{c} v^{c} \right)- \mathcal{C} \Big( \Omega^{I}{}_{J}
R^{J} \nonumber
\\
&& ~ - 2 e^{c}_{J} H^{c'}{}_{c} v^{d} \sigma^{b}{}_{dc'} e^{[I}_{b}
R^{J]} \Big) \Big] ~.
\end{eqnarray}
This uses the notation $j^{I}:=e^{I}_{a} j^{a}$,
$\mathcal{A}^{I}:=e^{I}_{a} \mathcal{A}^{a}$, etc. But it can be
rewritten more suggestively using (\ref{RhoDiverge}):
\begin{eqnarray}
j^{I} &=& \Delta N^{-1} \Big\{ \mathcal{A}^{I} + R^{I}
\dot{\mathcal{C}} - \left( \rho/\Delta \right) \left(
\Omega^{I}{}_{J} R^{J} - e^{I}_{a} v^{b} \sigma^{a}{}_{b} \right)
\nonumber
\\
&& ~+ 2 \partial_{J} \Big[ R^{[I} \Big( \Omega^{J]}{}_{K}R^{K}-
e^{J]}_{a} v^{b} \sigma^{a}{}_{b} \Big) \mathcal{C} \Big] \Big\} ~.
\end{eqnarray}
This form is useful because the second line is manifestly (three-)
divergence-free. From (\ref{divA}), $\mathcal{A}^{I}$ also has this
property, but is otherwise arbitrary (ignoring its support
properties for the moment). We may therefore absorb these terms by
defining
\begin{eqnarray}
\mathcal{J}^{I} &:=& \mathcal{A}^{I} + q R^{I} \dot{\psi} + 2
\partial_{J} \Big[ R^{[I} \Big( \Omega^{J]}{}_{K}R^{K} \nonumber
\\
&& ~ - e^{J]}_{a} v^{b} \sigma^{a}{}_{b} \Big) \mathcal{C} \Big] ~.
\label{LDefine}
\end{eqnarray}
The three-current then reduces to
\begin{eqnarray}
j^{I} &=& N^{-1} \Big[ \Delta \left( \mathcal{J}^{I} - R^{I}
\dot{\varphi} \right) + \rho \Big( e^{I}_{a} v^{b} \sigma^{a}{}_{b}
\nonumber
\\
&& ~ - \Omega^{I}{}_{J} R^{J} \Big) \Big] ~.
\label{jFinal}
\end{eqnarray}

In summary, a conserved current vector may be constructed by
choosing ``moment potentials'' $\varphi(\mathbf{R},z)$ and
$\mathcal{J}^{I}(\mathbf{R},z)$. $\varphi$ must reduce to $q \psi$
outside of $W$, but is otherwise arbitrary if we don't worry about
smoothness of $J^{a'}$. Similarly, $\mathcal{J}^{I}$ must reduce to
$q R^{I} \dot{\psi}$ outside of the worldtube. It is also required
to satisfy $\partial_{I} \mathcal{J}^{I}=0$ everywhere. Given
functions with these properties, the current is straightforward to
compute from (\ref{RhoFinal}), (\ref{jFinal}), and
\begin{equation}
J^{a'} = H^{a'}{}_{a} \left( n^{a} \rho + e^{a}_{I} j^{I} \right) ~.
\end{equation}
This last equation is a direct consequence of (\ref{rhoDefine}) and (\ref{jDefine}).

\section{Choosing $\varphi$ and $\mathcal{J}^{I}$}
\label{ChooseCurrent}

We saw in the previous section that nearly any current may be
constructed from potentials $\varphi$ and $\mathcal{J}^{I}$ chosen
to satisfy very simple algebraic matching conditions. But many such
choices generate singular current vectors. In this section, we show
how to exclude this class; allowing one to easily construct
physically reasonable divergence-free vector fields.

Since $\varphi$ must reduce to $q \psi$ outside of $W$, our first
step is to explicitly solve (\ref{PsiDefine}). Adopting spherical
coordinates $(r,\theta,\phi)$ defined in terms of $\mathbf{R}$ in
the standard way, the solutions of this equation are easily seen to
be of the form
\begin{equation}
\psi = \frac{1}{4\pi r^{3}} \left( 1 + q^{-1} A(\theta,\phi;z)
\right) ~, \label{psiSoln}
\end{equation}
where $A(\theta,\phi;z)$ is any function with a vanishing monopole
moment (in the elementary sense). The factor $q^{-1}$ is for later
convenience.

The arbitrariness of $A$ simply means that $\psi$ is not uniquely
determined by (\ref{PsiDefine}), which is to be expected. This
ambiguity translates into a freedom in $\varphi$ as well. The same
conclusion may also be reached directly from (\ref{RhoFinal}), where
a given $\rho$ only determines $\varphi$ up to terms of the form
$A/r^{3}$. But there cannot be a $1/r^{3}$ divergence in $\varphi$
(without any angular dependence), as this would imply that the
charge density contained an unphysical $\delta$-function at the
origin. It follows that there always exists a $\psi$ which
completely removes any $o(r^{-3})$ singularities in $\varphi$. From
now on, we always choose potentials in this class. This choice
actually removes the only polynomial singularity that may exist in
$\varphi$. Others would lead to unphysical charge densities.

Most $\delta$-function singularities are excluded from $\varphi$ on
similar grounds. The only exceptions are terms of the form $a(z)
\delta^{3}(\mathbf{R})$. But these do not affect the current or the
moments in any way, so they have no physical relevance. We can
therefore set $a(z)=0$ without any loss of generality.

Now choose a $\varphi$ that is continuous in both $\mathbf{R}$ and
$z$. As long as the boundary of the worldtube can be written in the
form $r=\mathcal{R}(\theta,\phi;z)$, where $\mathcal{R}$ is
single-valued, specifying $\varphi$ in some neighborhood of $W$
determines $\psi$ everywhere:
\begin{equation}
q \psi = \left( \frac{ \mathcal{R} }{ r } \right)^{3}
\varphi(\mathcal{R},\theta,\phi;z) ~. \label{PsiFromPhi}
\end{equation}
The problem with this approach is that arbitrary choices of
$\varphi$ will often lead to charges that vary over time.


It is much simpler if we look for a $\varphi$ that fits a given
$\psi$ rather than a $\psi$ which fits a given $\varphi$. This has
the advantage of allowing one to specify the charge \textit{a
priori}. It is then trivial to ensure that it remains constant.

Before showing how to do this, we first require that the charge
density at $r=0$ does not depend on the angular variables (i.e. that
it is single-valued). Define a central density $\rho_{0}(z)$ such
that
\begin{equation}
\rho_{0}(z) := \rho(0,\theta,\phi;z) ~.
\end{equation}

Now in order to find $\varphi$, choose $q$, $A$, $\rho_{0}$, and
$\mathcal{R}(\theta,\phi;z)$. For $r \leq \mathcal{R}$, $\varphi$ may then be taken to have the form
\begin{eqnarray}
\varphi &=& \frac{r}{\mathcal{R}} \left( \frac{q + A}{4\pi
\mathcal{R}^{3}} \right) + (1-r/\mathcal{R}) \nonumber
\\
&& ~ \times \left( \frac{1}{3} \rho_{0} + \frac{r}{\mathcal{R}}
\bar{\varphi} \right) ~. \label{PhiChoice1}
\end{eqnarray}
$\bar{\varphi}(r,\theta,\phi;z)$ is any bounded and continuous
function in $W$. As an example, a charge density of $\rho_{0}
\Delta$ is generated by choosing $\bar{\varphi}=0$ and
\begin{equation}
A = \frac{4 \pi}{3} \mathcal{R}^{3} \rho_{0} - q ~.
\end{equation}
One sees that $A$ vanishes in this case when the boundary is a sphere.

With very little extra effort, we could also take into account the
charge density on the body's surface, or any number of other
boundary conditions. This isn't particularly necessary, however, and
the extensions are straightforward.

The remainder of the current vector is now found by specifying an
arbitrary divergence-free ``three-vector'' $\mathcal{J}^{I}
(\mathbf{R},z)$ which reduces to $qR^{I} \dot{\psi}$ outside of the
body. If we choose a nonrotating triad, this vector field has a fairly straightforward physical interpretation in many cases of physical interest. Suppose that $n^{a}$ and $v^{a}$ are determined by center-of-mass conditions, and that the forces and torques on the body are not too large. Then $n^{a} \simeq v^{a}$, and the components of $\dot{n}^{I}$ would be small relative to the (reciprocal of the) worldtube's proper diameter. If the spacetime curvature inside $W$ is also small in an appropriate sense, (\ref{jFinal}) will reduce to $j^{I} \simeq \mathcal{J}^{I} - R^{I} \dot{\varphi}$. So physically, $\mathcal{J}^{I}$ may be interpreted as essentially that portion of the 3-current which arises due to the bulk motion ``internal'' charges. In this approximation, $\mathcal{J}^{I}$ should therefore be nonsingular throughout the body. A slightly more detailed look at (\ref{jFinal}) and (\ref{PhiChoice1}) shows that this observation must remain true even in the exact theory.

We are therefore left with the problem of writing down a nonsingular
divergence-free three-vector which has a specified form outside of a
given boundary. Finding vector fields with these properties may be
nontrivial when $\dot{A} \neq 0$, so we now illustrate a method
that may simplify the process. First write down $\mathcal{J}^{I}$ so
that it automatically satisfies the appropriate boundary condition:
\begin{equation}
\mathcal{J}^{I} = \Theta(\mathcal{R}-r) \mathcal{J}^{I}_{(-)} +
\Theta(r-\mathcal{R}) q R^{I} \dot{\psi}  ~. \label{Jansatz}
\end{equation}
Here, $\Theta(\cdot)$ is the Heaviside step function; $\Theta(x) = 1$ if $x>0$, but otherwise vanishes.

This expression also introduces the new vector field $\mathcal{J}^{I}_{(-)}
(r,\theta,\phi;z)$ which must be both nonsingular and
divergence-free. While we could require this to exactly match $q
R^{I} \dot{\psi}$ at $r=\mathcal{R}$, doing so would eliminate the
possibility of currents flowing on the surface of the body.
Moreover, the divergence of (\ref{Jansatz}) will vanish everywhere
if only the normal component of $\mathcal{J}^{I}$ is continuous
across the boundary. We therefore suppose only that
\begin{equation}
\left( s \cdot \mathcal{J}_{(-)} \right)_{r=\mathcal{R}} = \frac{
\dot{A} }{4 \pi R^{2}} ~, \label{Bdry1}
\end{equation}
where
\begin{equation}
s_{I} := \partial_{I} \left( r - \mathcal{R} \right) ~.
\end{equation}

The boundary condition (\ref{Bdry1}) is difficult to apply directly
if we (say) write $\mathcal{J}^{I}_{(-)}$ in terms of a vector
potential. To fix this, we instead construct it from two scalar
potentials $Q$ and $P$ in such a way that its divergence will always
vanish. To simplify the notation, it is first useful to define the
operator
\begin{equation}
\mathbb{L}_{I} := \epsilon_{IJK} R^{J} \partial^{K} ~,
\end{equation}
or ``$\mathbb{L} := \mathbf{R} \bm{\times} \nabla$'' in a more traditional notation. This is just ($i/\hbar$ multiplied by) the standard angular momentum operator used in quantum mechanics.

We may now write
\begin{equation}
\mathcal{J}^{I}_{(-)} =  \epsilon^{IJK} \partial_{J} \mathbb{L}_{K}
Q + \mathbb{L}^{I} P ~. \label{VectPotential}
\end{equation}
This expansion is essentially an application of the angular momentum
Helmholtz theorem \cite{Moses}. The scalars may also be interpreted
as Debye potentials \cite{Nisbet, Scalars}, or the generators of the
toroidal and poloidal components of $\mathcal{J}^{I}_{(-)}$.

The utility of this expansion can be seen by considering scalar
potentials which may be split into angular and (normalized) radial
components. For example, let
\begin{eqnarray}
Q(r,\theta,\phi;z) &=& \sum_{\ell=0}^{\infty}
\bar{Q}_{\ell}(r/\mathcal{R};z) \hat{Q}_{\ell}(\theta,\phi;z) ~,
\label{EllSplit}
\end{eqnarray}
and break up $P$ in the same way. We choose the first term to
generate the right-hand side of (\ref{Bdry1}). For $\ell \geq 1$,
the angular functions $\hat{Q}_{\ell}$ could then be taken as
appropriate basis functions. The corresponding $\bar{Q}_{\ell}$ will
be chosen to ensure that the $\ell$th term in the sum doesn't add
anything to the normal component of $\mathcal{J}^{I}_{(-)}$ at the
boundary.

Looking at the $\ell=0$ terms, (\ref{Bdry1}) becomes
\begin{eqnarray}
\frac{\dot{A}}{4 \pi} &=& \mathbb{L} \cdot \Big[ \bar{Q}_{0}
\mathcal{R} \mathbb{L} \hat{Q}_{0} - \bar{Q}_{0}' \hat{Q}_{0}
\mathbb{L} \mathcal{R} \nonumber
\\
&& ~ - \bar{P}_{0} \hat{P}_{0} \mathcal{R}^{2} \nabla \mathcal{R}
\Big] ~, \label{Bdry2}
\end{eqnarray}
where $\bar{Q}_{0}'$ indicates a derivative of $\bar{Q}_{0}$ with
respect to $r/\mathcal{R}$. Also note that these radial functions
are understood to already be evaluated at $r/\mathcal{R}=1$ (i.e.
$\mathbb{L}$ does not act on them here).

The solutions to (\ref{Bdry2}) are clearly not unique (even taking
into account the gauge freedom). Still, we only need to generate
\textit{a} solution to this equation. Supposing that
$\bar{P}_{0}=\hat{P}_{0}=0$, $\bar{Q}_{0}(1;z)=1$, and
$\bar{Q}'_{0}(1;z)=-1$, we are left with the relatively simple
result
\begin{equation}
\frac{\dot{A}}{4\pi} = \mathbb{L}^{2} \left( \mathcal{R} \hat{Q}_{0}
\right) ~.
\end{equation}
This may be solved by expanding $\mathcal{R} \hat{Q}_{0}$ in
spherical harmonics, using the addition theorem to express the
series in terms of Legendre polynomials, and then summing the
result. One eventually finds that (up to a possible additive
function proportional to $1/\mathcal{R}$)
\begin{eqnarray}
\hat{Q}_{0}(\theta,\phi;z) &=& - \frac{1}{16 \pi^{2} \mathcal{R}}
\int \mathrm{d} \Omega' \dot{A}(\theta',\phi';z) \nonumber
\\
&& ~ \times \ln\left(1-\cos \alpha \right) ~. \label{AngEll}
\end{eqnarray}
$\mathrm{d} \Omega'$ is that standard volume element on a sphere,
and $\alpha$ is the angle between the point of integration and
the point where $\hat{Q}_{0}$ is being evaluated.

The radial profile $\bar{Q}_{0}$ may now be chosen. This must
satisfy the aforementioned conditions $\bar{Q}_{0}(1;z)=1$ and
$\bar{Q}_{0}'(1;z)=-1$, but we must also require
\begin{equation}
\lim_{r \rightarrow 0} \bar{Q}_{0}(r/\mathcal{R};z)/r = 0 ~.
\end{equation}
This is necessary to ensure that the three-current is nonsingular
and single-valued at the origin. A particularly simple choice is
then
\begin{eqnarray}
\bar{Q}_{0}(r/\mathcal{R};z) &=& \left( \frac{r}{\mathcal{R}}
\right)^{2} \left( 4- 3 \frac{r}{\mathcal{R}} \right) ~.
\label{RadEll}
\end{eqnarray}
This completes our solution to (\ref{Bdry2}).

But as was already mentioned, the full solution to (\ref{Bdry1}) is usually more complicated. If each term in (\ref{EllSplit}) may be considered individually, then for all $\ell \geq 1$,
\begin{eqnarray}
\bar{Q}_{\ell} \mathcal{R} \mathbb{L}_{I} \hat{Q}_{\ell} &=&
\bar{Q}'_{\ell} \hat{Q}_{\ell} \mathbb{L}_{I} \mathcal{R}+
\bar{P}_{\ell} \hat{P}_{\ell} \mathcal{R}^{2} \partial_{I} \mathcal{R}
\label{Bdry3}
\end{eqnarray}
at $r=\mathcal{R}$.

If $\partial_{I} \mathcal{R}=0$, this implies that either
$\bar{Q}_{\ell}(1;z)$ vanishes, or that $\hat{Q}_{\ell}$ is independent
of $\theta$ and $\phi$. But throughout the body, only the angular
derivatives of $\hat{Q}_{\ell}$ have any physical relevance, so this case is equivalent to $\hat{Q}_{\ell} = 0$. But this is itself indistinguishable from saying that $\bar{Q}$ vanishes everywhere. So we
may always say that $\bar{Q}_{\ell}$ vanishes at $r=\mathcal{R}$ when the boundary is locally spherical.

In this case, $\mathcal{J}^{I}_{(-)}$ reduces to a sum of terms of the form
\begin{equation}
\bar{Q}_{\ell} \left( \frac{R^{I}}{r^{2}} \mathbb{L}^{2} \hat{Q}_{\ell}
- \partial^{I} \hat{Q}_{\ell} \right) + \frac{r}{\mathcal{R}}
\bar{Q}'_{\ell} \partial^{I} \hat{Q}_{\ell} + \bar{P}_{\ell} \mathbb{L}^{I} \hat{P}_{\ell} ~. \label{IntCurrent}
\end{equation}
On the boundary, the first group of terms clearly vanishes for all $\ell
\geq 1$.

But we must also make sure that (\ref{IntCurrent}) is everywhere
finite. We first require none of the functions $\bar{P}_{\ell}$ contain any singularities. Choosing $\bar{Q}$ is slightly more complicated due to the $o(r^{-1})$ divergence in (\ref{IntCurrent}). But for all $\ell \neq 0$, we may suppose that there exists a nonsingular function $Q_{\mathrm{r}, \ell}$ such that
\begin{equation}
\bar{Q}_{\ell} (r/\mathcal{R};z) = \frac{r}{\mathcal{R}} \left( 1 -
\frac{r}{\mathcal{R}} \right) Q_{\mathrm{r}, \ell}(r/\mathcal{R};z)
~. \label{RadProfile}
\end{equation}
This guarantees that (\ref{IntCurrent}) remains finite. As long as the sum over $\ell$ does not diverge, these choices are sufficient to describe a wide variety of functions $\mathcal{J}^{I}$ (for angles) where $\partial_{I} \mathcal{R} = 0$.

If the boundary radius is changing, the situation is more complicated,
and our discussion will not be quite so complete. The full equation
(\ref{Bdry3}) must be solved in general at $r=\mathcal{R}$, although
it isn't obvious how to do this. To make some progress, suppose that
$\bar{P}_{\ell}$ vanishes on the boundary. If $\bar{Q}_{\ell}(1;z)$
also vanishes, $\hat{Q}_{\ell}$ may be anything at all (so long as it
remains finite). But more generically, it must have the form
\begin{equation}
\hat{Q}_{\ell} \propto \mathcal{R}^{\left[
\bar{Q}'_{\ell}(1;z) / \bar{Q}_{\ell}(1;z) \right] } ~.
\end{equation}

The general case where $P_{\ell} \neq 0$ at $r=\mathcal{R}$ does not
seem simple to discuss. It is likely that an ansatz better
adapted to nonspherical boundaries than (\ref{VectPotential}) would
be useful for this purpose. The class of currents which we are
excluding by neglecting this case does not really seem significant,
however.

To summarize, $\mathcal{J}^{I}$ is required to compute the
three-current. It has the form (\ref{Jansatz}), where
$\mathcal{J}^{I}_{(-)}$ may be expanded using (\ref{VectPotential}).
The scalar potentials in this expression have the form
(\ref{EllSplit}). By definition, $P_{0}=0$, and $\hat{Q}_{0}$ and
$\bar{Q}_{0}$ are given by (\ref{AngEll}) and (\ref{RadEll})
respectively. The higher-$\ell$ components of the potentials may be
chosen by first writing down any (smooth) sets of functions
$\bar{P}_{\ell}$, $\hat{P}_{\ell}$, and $\hat{Q}_{\ell}$. Then each
$\bar{Q}_{\ell}$ should have the form (\ref{RadProfile}). Note that
while these choices do not allow for every possibility, they are
still quite general.

In all of this, the $z$-dependence of $\varphi$ and
$\mathcal{J}^{I}$ is essentially arbitrary. Since $z(s)$ is a
one-to-one function, these potentials may easily be rewritten as
functions of $(\mathbf{R},s)$ rather than $(\mathbf{R},z)$. When
doing so, it seems sufficient to require that they be $C^{1}$ in
$s$. All of these conditions taken together allow us to easily
generate a very wide class of physically reasonable current vectors with a given $q$. Note that the continuity restrictions are sufficient but not necessary. Still, the stated classes seem large enough for almost
any practical purposes.

Now that acceptable classes of $\varphi$ and $\mathcal{J}^{I}$ have been identified, expressions in Sec. \ref{ConstructCurrent} and appendix \ref{DixReview} may be used to find the which sets of moments correspond to plausible current vectors. Since translating between these two representations involves a Fourier transform, it is likely to be difficult to do this explicitly. We shall not attempt it, but merely state that most of the restrictions on the moments would involve their asymptotic behavior in the ``large-$n$'' limit.

\section{Dynamical rigidity}
\label{Rigidity}

In Newtonian mechanics, one often makes use of the concept of rigid
bodies. Although such objects do not (and cannot) exist, the idea is
still useful. In many cases, the bulk dynamics of a large class of
real materials are known to be adequately approximated by the
behavior of a suitable rigid body. Just as importantly, rigidity
conditions allow equations of motion to be solved uniquely in
relatively simple ways. Although it is not at all clear that a
similarly useful concept should exist in highly relativistic
systems, it is still interesting to search for one that is roughly
analogous.

Various definitions of rigidity in curved spacetimes have been
introduced over the years. But perhaps the simplest is a generalization
of the concept advanced by Born for objects in Minkowski space
\cite{Born, Born2, Born3}. This essentially requires that the strain
tensor throughout a material be Lie-dragged along the worldlines of
its constituent particles. Unfortunately, such a restriction fails
to be self-consistent in many systems of practical interest
\cite{NoBorn}.

Rather than defining rigidity through local kinematics, it has been
conjectured that fixing certain ``global'' (or quasilocal)
quantities may be more fruitful. In particular, one may wish to call
a body rigid if its multipole moments evaluated in some particularly
natural frame do not change.

Supposing again that we are discussing the mechanics of some extended body, the reference line $Z$ and the dynamical velocity $n^{a}$ of this frame may be chosen using the center-of-mass conditions given in \cite{EhlRud, Dix79} (and rigorously justified in \cite{CM}). In doing so, the temporal component of the tetrad would be proportional to the body's bulk (linear) momentum vector. The triad vectors would not be chosen uniquely in this scheme, although their evolution may be fixed by defining an ``average angular velocity'' as the vector which must multiply the ``inertia tensor'' to yield the bulk angular momentum vector  \cite{EhlRud, Dix79}. This effectively defines $\Omega^{I}{}_{J}$ in terms of the body's stress-energy tensor.

These ideas were introduced in \cite{EhlRud} under the name of
quasirigidity, and supposed that the multipole moments under
discussion were those of the object's stress-energy tensor. Very
similar ideas have also been referred to as dynamical rigidity in
\cite{Dix70a, Dix79}, and we shall use the two terms
interchangeably. These latter papers also introduced a scalar defined from the body's stress-energy tensor (and the metric) which may be interpreted as its total internal energy. It was found that this is in fact a
constant of motion for bodies which are dynamically rigid. The
concept therefore has an elegant physical interpretation.

Changing definitions slightly, the results of the previous sections
may be used to explicitly construct quasirigid bodies. We do so by
requiring that the multipole moments of the number density and
electric current vectors associated with a given object be fixed in
the aforementioned frame (rather than the moments of its
stress-energy tensor). This is actually far less restrictive than
the definition in \cite{EhlRud, Dix79}, as it ignores the stress and
energy density distributions inside the body. In doing so, it also
appears closer in spirit to the concept of Newtonian rigidity. A
downside is that the total internal energy of the body will no
longer be conserved in general. It is also somewhat less ``dynamical''
than the original definition.

There is one potential ambiguity in this: it might be useful to
allow the monopole moment to vary even if the higher moments do not.
If $Q^{A} := e_{a}^{A} Q^{a}$ were held constant, (\ref{Monopole}) implies that $q
\partial (v^{I}) / \partial s = 0$, or
\begin{equation}
q \left( \dot{n}^{I} - \dot{v}^{I} - \Omega^{I}{}_{J} v^{J} \right)
=0  ~. \label{RigidMonopole}
\end{equation}
This is of course satisfied trivially if either $q=0$ or
$n^{a}=v^{a}$ and $\Omega^{I}{}_{J}=0$. But in general, it provides
a rather strict restriction on the class of allowed frames.

In a mechanics problem, one would usually derive $v^{a}$ from
$n^{a}$ using the mass, force, torque, and linear and angular
momenta of the body \cite{EhlRud, Dix79}. In these cases,
(\ref{RigidMonopole}) would restrict which combinations of these
objects could be allowed, which seems unphysical. In any case, one
of the main reasons for adopting rigidity conditions in the first
place is to obtain unique evolution equations for all relevant
quantities. But changes in $Q^{a}$ would already be fixed in most
cases of interest. There is no need to force a potentially
contradictory condition on top of the already existing one. We
therefore define a current to be quasirigid if all of its moments
except for (possibly) the monopole are constant.

It is clear from (\ref{MomentDefine}) that this definition implies
that $\check{J}^{a}(X,z) e_{a}^{A}(z) = \check{J}^{A}$ is a function
purely of $e^{A}_{a} X^{a} = R^{A}$. Noting (\ref{Jcheck}), this
means that $\mathcal{A}^{I}$ and $\mathcal{B}$ must be independent
of $z$ (or equivalently, of $s$).

We first consider the effect of holding $\mathcal{B}$ constant.
Supposing that $\rho$ is at least instantaneously smooth,
(\ref{CDefine}) and (\ref{RhoDiverge}) can be used to show that
\begin{equation}
\dot{\rho} \sim \Delta \left( \mathcal{B}/ \tilde{N} \right)  \frac{
\partial }{ \partial s} \left( R^{I}
\partial_{I} \ln \tilde{N} \right) ~,
\end{equation}
where terms that are obviously nonsingular have been suppressed.

In general, $\mathcal{B}$ diverges at the origin. The charge density
may therefore develop a singularity of its own if we are not
careful. One necessary (but not sufficient) condition that must be
satisfied to avoid this is
\begin{equation}
\left( q + A \right) \lim_{r \rightarrow 0} \frac{
\partial }{ \partial s} \left[ \left( R^{I}/r \right)
\partial_{I} \ln \tilde{N} \right] = 0 ~.
\label{Sing1}
\end{equation}

Writing this in a more useful form requires an expansion of
$\tilde{N}$ in powers of $\mathbf{R}$ or $X$ around $r=0$. But that in turn requires a similar expansion for $\sigma^{a}{}_{b}$. Such an expression may be found in \cite{PoissonRev}:
\begin{equation}
\sigma^{a}{}_{b} ( X,z ) = \delta^{a}_{b} - \frac{1}{3}
R^{a}{}_{cbd}(z) X^{c} X^{d} + \ldots ~,
\end{equation}
Here, $R^{a}{}_{cbd}(z)$ is the Riemann tensor at the origin. Using
this in (\ref{LapseDefine}), we find that the flattened lapse has an expansion of the form
\begin{equation}
\tilde{N} \simeq 1 + \dot{n}_{I} R^{I} + \frac{1}{3} \big( R_{0I0J} + R_{0IKJ} v^{K} \big) R^{I} R^{J} ~. \label{LapseExpand}
\end{equation}
This allows (\ref{Sing1}) to be reduced to
\begin{equation}
\left( q + A \right) \lim_{r \rightarrow 0} \left( R^{I}/r \right)
\left( \ddot{n}_{I} + \Omega_{I}{}^{J} \dot{n}_{J} \right) = 0 ~.
\end{equation}

There are two ways to satisfy this equation: either $q=A=0$ (i.e. $q \psi
=0$), or
\begin{equation}
\frac{\partial (\dot{n}_{I})}{\partial s} = \ddot{n}_{I} +
\Omega_{I}{}^{J} \dot{n}_{J} = 0~. \label{RigidCondition1}
\end{equation}

(\ref{Sing1}) isn't the only condition that must be satisfied in
order for $\mathcal{B}$ to remain constant. One must also have
\begin{equation}
\left( q + A \right) \lim_{r \rightarrow 0} \frac{
\partial }{ \partial s} \left[ \left( R^{I}/r^{2} \right)
\partial_{I} \ln \tilde{N} \right] = 0 ~.
\label{Sing2}
\end{equation}
If $q \psi = 0$, this is trivially satisfied. Otherwise
(\ref{RigidCondition1}) must hold, in which case (\ref{Sing2})
reduces to
\begin{equation}
\frac{\partial }{ \partial s} \left( R_{0I0J} + R_{0(I|K|J)} v^{K}
\right) = 0 ~. \label{RigidCondition2}
\end{equation}
Roughly speaking, (\ref{RigidCondition1}) and (\ref{RigidCondition2})
imply that the inertial and tidal forces felt by a center-of-mass
observer must remain constant.

This leads to a very important conclusion: physically acceptable
charged currents cannot be dynamically rigid with respect to arbitrary reference frames. Unless the system is particularly simple, no charged rigid current can exist. Although the structure of $J^{a'}$ will (usually)
affect $\dot{n}^{a}$ and $R^{a}{}_{bcd}(z)$, it seems very unlikely
that quasirigidity alone could conspire to ensure that they satisfy
(\ref{RigidCondition1}) and (\ref{RigidCondition2}) except in a few
special cases.

One consequence of this is that the only quasirigid particle fluxes
that could exist with generic applied forces must contain zero total
particles. Nontrivial particle fluxes therefore cannot be rigid
except for special types of motion.

Interesting rigid currents of electric charge always exist, however.
This is because although $q$ must still vanish in general, the charge density may change sign and $J^{a'}$ needn't be future-directed timelike.
Characterizing such vector fields is relatively simple. We start by
making sure that $\mathcal{B}$ is independent of $s$. Again, setting
$q \psi =0$ is sufficient. Then
\begin{equation}
\varphi = - \mathcal{B} / \tilde{N} ~. \label{RigidPhi}
\end{equation}
It is obvious from (\ref{PhiChoice1}) that there always exists a
large class of $\mathcal{B}$'s for which this is possible. Note,
however, that they all vanish at the edge of the body. This implies
that $\mathcal{R}$ cannot change (except perhaps if $\mathcal{B} =
\varphi = 0$). It is easy to verify that $\rho_{0}$ also remains
constant.

Besides requiring that $\dot{\mathcal{B}}=0$, quasirigidity also demands
that $\partial (\mathcal{A}^{I})/\partial s$. Suppose again that
$q\psi$ vanishes, so $\varphi$ satisfies (\ref{RigidPhi}). According
to (\ref{LDefine}), (\ref{jFinal}), and (\ref{psiSoln}), the
evolution equation for the three-current never become singular.
There therefore don't appear to be any particularly interesting
restrictions arising from the constancy of $\mathcal{A}^{I}$.

Note however that if $\dot{\mathcal{R}}=0$ was not already ensured
by $\dot{\mathcal{B}}=0$, then it is implied by the
time-independence of $\mathcal{A}^{I}$. Although rigidity requires
that the boundary of the object and its central charge density stay
constant in the pseudo-Fermi coordinate system we have been using,
$\rho$ and $j^{I}$ will still depend on $s$ in most cases. Their
evolution equations are rather lengthy, but easily derived from this
discussion (supplemented with some straightforward bitensor
identities).

To reiterate, we have shown that charged quasirigid currents often
do not exist. As with Born's original concept of relativistic
rigidity, the idea of unchanging multipole moments has significant
limitations. Of course, the definitions we've adopted are rather
strict. Although (\ref{RigidCondition1}) and (\ref{RigidCondition2})
can rarely be satisfied exactly, their left-hand sides will often be
very small in realistic physical systems. When this is true,
currents may generically exist that are in some sense ``nearly
rigid.'' But even when viewed from afar, the mechanical properties
of such objects are likely to be quite different from those of
``equivalent'' rigid bodies (which must contain singularities). In
particular, the computation of gravitational or electromagnetic
self-forces in a mechanical system would likely be problematic.

Our definition of dynamical rigidity could also be argued to be
overly strict in that it requires all of the current's multipole
moments to be fixed. This has the advantage of producing unique
evolution equations, but it isn't very well-motivated physically. It
would usually be more realistic to suppose that only the first few
moments are constant (or approximately constant). This is  closer in
spirit to the type of quasirigidity discussed in (for example)
\cite{LRigid2}, which was shown to have many desirable
properties. There does not appear to be any existence problem for
objects with only ``low order'' quasirigidity. But since the
evolution of the higher moments is not constrained by such a scheme,
the motion of the current cannot be unique.

It should also be stressed that the
definition of dynamical rigidity adopted here is different from the
one given in \cite{EhlRud, Dix79}. There, the constants of motion
were multipole moments of the stress-energy tensor rather than
moments of a relevant current vector. This condition appears more
restrictive, so we conjecture that a similar non-existence result
can also be proven for that case. It would then follow that the
total internal energy defined in \cite{Dix79} could never be
constant except in particularly simple systems.

\section{Comparable currents}
\label{Comparisons}

It is often useful to discuss the properties of objects in a highly dynamical spacetime with language borrowed from something simpler (such as Kerr or Minkowski). For example, one might want to say that a neutron star in some complicated numerical simulation was essentially the same star that had already been studied in isolation. Although defining ``same'' in this sentence may be quite tricky, there are several ways to proceed depending on the focus of the problem and the amount of effort one is willing to expend on it. This section will introduce one method which is particularly simple from a mathematical perspective, and then study some of its physical consequences.


We define two currents (possibly in different spacetimes) to be ``identical'' if their multipole moments evaluated in some pair of frames are identical. Distinguishing quantities related to the second current with underlines, we also require these frames to satisfy $\dot{n}^{I}(s) = \underline{\dot{n}}^{I}(s)$, $v^{I} = \underline{v}^{I}$, and $\Omega^{I}{}_{J}(s) = \underline{\Omega}^{I}{}_{J}(s)$ for all $s$. The two currents should also have identical charges: $q =
\underline{q}$. These conditions automatically imply that the frame
components of the monopole moments of both currents are identical.

Equality of the higher moments implies that $\mathcal{B}(\mathbf{R};s) =
\underline{\mathcal{B}}( \mathbf{R} ;s)$ and $\mathcal{A}^{I}(
\mathbf{R};s ) = \underline{\mathcal{A}}^{I}( \mathbf{R}; s)$.
Finding the consequences of these relations will use arguments very
similar to those given in the previous section.

As a consequence of both $\mathcal{B}$'s being equal,
(\ref{CDefine}) and (\ref{RhoDiverge}) imply that
\begin{equation}
\mathcal{B} R^{I} \partial_{I} \ln \tilde{N}/\underline{\tilde{N}} =
\left( q - \underline{\rho} \right) \underline{\tilde{N}} /
\underline{\Delta} - \left( q - \rho \right) \tilde{N} / \Delta~.
\label{BComp}
\end{equation}
In physically reasonable systems, the right-hand side of this
equation is clearly nonsingular. But $\mathcal{B}$ generally
involves an $o(r^{-3})$ singularity. This divergence is softened by the fact that $R^{I} \partial_{I} \ln \tilde{N} / \underline{\tilde{N}}$ decreases at least as fast as $r^{2}$ near the origin. But this is not enough to allow $\mathcal{B}$ to be chosen arbitrarily (using only the constraints discussed thus far). In order to remain consistent, it must be true that
\begin{equation}
\left( q + A \right) \lim_{r \rightarrow 0 } \left[ \left( R^{I} /
r^{2} \right) \partial_{I} \ln \tilde{N} / \underline{\tilde{N}}
\right] = 0 ~.
\end{equation}
Using (\ref{LapseExpand}), this is equivalent to either $q \psi = 0$
or
\begin{equation}
R_{0I0J}+R_{0(I|K|J)} v^{K} = \underline{R}_{0I0J} +
\underline{R}_{0(I|K|J)} v^{K} ~. \label{CompCondition}
\end{equation}
As before, these Riemann tensors are evaluated at the respective
origins of each frame. This equation essentially states that
physically reasonable charged currents can only be comparable if
observers attached to the center of each frame would experience
identical tidal forces. If two charged currents were ``comparable''
(as defined above), but did not satisfy (\ref{CompCondition}), at
least one must be singular.

We may now construct comparable currents satisfying $q \psi =0$. In
this case,
\begin{equation}
\varphi = \left( \underline{\tilde{N}} / \tilde{N} \right)
\underline{\varphi} ~.
\end{equation}
It follows that $\rho_{0} = \underline{\rho}_{0}$ and $\mathcal{R} =
\underline{\mathcal{R}}$ (at least if $\mathcal{B} \neq 0$).

Continuing, $\mathcal{A}^{I}$ must be equal to
$\underline{\mathcal{A}}^{I}$ if two current are to be comparable.
Suppose again that $q \psi=0$. If $\mathcal{B}$ also vanishes, the
two current potentials would become equal: $\mathcal{J}^{I} =
\underline{\mathcal{J}}^{I}$. This would imply that the frame
components of the two three-currents would be be directly
proportional to one another:
\begin{equation}
j^{I} = \left( \frac{ \Delta N^{-1} }{ \underline{\Delta}
\underline{N}^{-1} } \right) \underline{j}^{I} ~.
\end{equation}

In any case, the situation here is similar to the one obtained in the previous section. The idea of having two objects where all of the relevant moments are identical can only be generic if they both carry zero charge.

\section{Discussion}

We have shown how to construct a vector field $J^{a'}$ satisfying
(\ref{DivFree}) from a set of multipole moments defined with respect
to a given reference frame. In so doing, we have implicitly found
restrictions on the moments which ensure that the resulting current
will be smooth and bounded. This effectively completes the program
Dixon initiated in \cite{Dix67, Dix70b} to study the multipole
moments of nonsingular conserved vector fields in curved spacetime.

An immediate application of these constructions is a method to
easily construct currents with a given total charge $q$. This is
described in detail in Sec. \ref{ChooseCurrent}, where a few free
functions with relatively intuitive interpretations are used to
build any smooth $J^{a'}$.

In combination with the equations of motion derived in \cite{Dix74}, one could (for example) derive the equations of motion for all charged particles with a given charge in some approximation scheme. This was carried out in \cite{Harte} in flat spacetime in part to determine precisely when the Lorentz-Dirac equation is valid. The results derived in this paper would then allow similar methods to be used to study the validity of its generalization, the Dewitt-Brehme equation
\cite{PoissonRev}.

The methods used here may also be seen as a prototype for a similar analysis of the multipole moments of the stress-energy tensor (as defined in \cite{Dix74, Dix79}). This would allow one to directly construct all bodies having prescribed linear and angular momenta at a given center-of-mass position.

\begin{acknowledgements}

The author wishes to thank Pablo Laguna for helpful discussions. This work was supported by NSF grant PHY-0555436. The CGWP is supported by the NSF under cooperative agreement PHY-0114375.

\end{acknowledgements}

\appendix

\section{Bitensors}
\label{Bitensors}

This appendix briefly reviews the theory of bitensors. The detailed
properties of such objects have been discussed in detail by several
authors \cite{Synge,PoissonRev, WaveCurved, Dix79}, so we shall
include only a few relevant definitions and identities.

Our discussion will focus on the world function biscalar
$\sigma$. Given two points $x$ and $y$ in a spacetime, it is first assumed that there exists exactly one geodesic connecting them. In this paper, these points will always be spacelike-separated, so this is a rather mild restriction. Regardless, $\sigma(x,y)$ is defined to be one half of the square of the geodesic distance between its arguments.

The derivatives of $\sigma$ arise in many expressions, so it is standard to define a special notation for them. In particular, let
\begin{eqnarray}
\sigma_{a}(x,z) &:=& \nabla_{a} \sigma(x,z)  ~,
\\
\sigma_{a'}(x,z) &:=& \nabla_{a'} \sigma(x,z)  ~.
\end{eqnarray}
Here, we have assumed that $z \in Z$, and used the
previously-mentioned convention that indices referring to points on
$Z$ are not primed. This notation trivially generalizes to
expressions involving any number of derivatives. For example,
$\sigma_{aba'c}=\nabla_{ca'ba}\sigma$. Derivatives at $x$ always
commute with derivatives at $z$, although the order of derivatives
at a particular point cannot generally be interchanged (as is usual
for covariant derivatives).

It may be shown that these vectors are tangent to the geodesic connecting $x$ and $z$, and that their magnitudes are just the squared geodesic distance \cite{Synge, PoissonRev,
WaveCurved, Dix79}:
\begin{equation}
\sigma^{a} \sigma_{a} = \sigma^{a'} \sigma_{a'} = 2 \sigma ~.
\label{MasterIdent}
\end{equation}
This allows one to naturally interpret $\sigma_{a}$ and $\sigma_{a'}$ as generalized displacement vectors between $x$ and $z$.

Differentiating (\ref{MasterIdent}) an appropriate number of times also allows one to generate a number of useful identities. The
simplest of these are $\sigma_{a} \sigma^{a}{}_{b} = \sigma_{b}$,
$\sigma_{a} \sigma^{a}{}_{a'} = \sigma_{a'}$, etc. It follows that
$\sigma^{a}{}_{b} = \delta^{a}_{b}$ in flat spacetime, or in general
when $x = z$. Similarly, $\sigma^{a}{}_{a'} = -\delta^{a}_{a'}$ in
these cases.

The matrix inverse of $-\sigma^{a}{}_{a'}$ appears a number of
times in Dixon's formalism, so it is convenient to define
\begin{equation}
H^{a'}{}_{a}(x,z) := \left[ - \sigma^{a}{}_{a'}(x,z) \right]^{-1} ~.
\label{HDefine}
\end{equation}
This object also appears naturally in solutions of the geodesic
deviation (Jacobi) and Killing transport equations \cite{Dix70a,
Dix79}. Its derivatives have the form
\begin{equation}
\nabla_{b'} H^{a'}{}_{a} = H^{a'}{}_{b} H^{c'}{}_{a}
\sigma^{b}{}_{b'c'} ~. \label{HDerivative}
\end{equation}

Another bitensor with similar properties is the parallel propagator
$g^{a'}{}_{a}(x,z)$. This does exactly what its name implies:
parallel propagates vectors from $z$ to $x$ along the geodesic
connecting them. The properties of this object may be found in
\cite{Synge,PoissonRev}. For our purposes, note that
$g_{a'a}(x,z)=g_{aa'}(z,x)$. From its physical interpretation, it
also follows that
\begin{eqnarray}
g^{a'}{}_{a} g^{a}{}_{b'} &=& \delta^{a'}_{b'} ~,
\\
g^{a}{}_{a'} g^{a'}{}_{b} &=& \delta^{a}_{b}  ~.
\end{eqnarray}

An important biscalar built from $g^{a'}{}_{a}$ is known as the van
Vleck determinant $\Delta(x,z)$. It is defined to be
\begin{equation}
\Delta(x,z) := \det\left( -g^{a'}{}_{a} \sigma^{a}{}_{b'} \right) ~.
\label{DeltaDefine}
\end{equation}
In \cite{PoissonRev}, it is shown that
\begin{equation}
\det\left( g^{a'}{}_{a} \right) = \sqrt{ \frac{ g(z) }{ g(x) } } ~,
\end{equation}
so
\begin{equation}
\Delta(x,z) =  \sqrt{ \frac{ g(z) }{ g(x) } } \det\left( -
\sigma^{a}{}_{a'} \right) ~. \label{vanVleckExpand}
\end{equation}
The van Vleck determinant plays a fundamental role in the focusing
of geodesic congruences \cite{PoissonRev, VanVleck}. But here, it
will appear as an important factor in the equations giving the
current vector in terms of quantities constructed from the moments.
It arises essentially from a coordinate transformation performed in
the derivation.

In particular, a number of relevant functions that we deal with are
defined on the tangent bundle $TM$ restricted to $Z$. Occasionally,
a base point $z$ will be fixed, in which case we can switch freely
between considering these objects to be functions on $TM$ (really
just the tangent space $T_{z}M$) or on the manifold itself (at least
in a neighborhood of $z$). This identification will be made via the
exponential map, so abusing the notation slightly, we make
identifications of the form $f(X,z) = f(\exp_{z} X)$ for any
function $f$ and vector $X \in T_{z}M$.

It is then useful to transform integrals over spacetime into
integrals over $T_{z}M$. The inverse of the exponential map is just
$\sigma^{a}$. So define
\begin{equation}
X^{a}(x,z) := - \sigma^{a}(x,z) ~.
\label{XDefine}
\end{equation}
This acts as a ``radial vector'' between $x$ and $z$, and from (\ref{vanVleckExpand}), the Jacobian of the transformation $x
\rightarrow X$ is
\begin{equation}
\left| \frac{\partial X}{\partial x} \right| = \sqrt{ \frac{ g(x) }{
g(z) } } \Delta ~. \label{Jacobian}
\end{equation}

Lastly, we note that directly differentiating (\ref{DeltaDefine})
yields
\begin{equation}
\nabla_{a} \ln \Delta = - H^{b'}{}_{b} \sigma^{b}{}_{ab'} ~.
\label{DeltaDerivative}
\end{equation}
Other properties of the van Vleck determinant may be found in
\cite{PoissonRev,VanVleck}.

\section{Dixon's formalism}
\label{DixReview}

In the present context, Dixon's formalism \cite{Dix67, Dix70a,
Dix70b, Dix74, Dix79, EhlRud} defines a set of multipole moments for
spatially-bounded vectors satisfying (\ref{DivFree}). This set is
complete in the sense that specifying all of the moments allows one
to completely reconstruct $\mathbf{J}$. If this vector field is
interpreted as the electromagnetic current of a charged test body,
it is possible to show that these moments also have the standard
interpretation in terms of expansions for the electromagnetic force
and torque in successively higher derivatives of the applied field
\cite{Dix67, Dix74}.

In most other formalisms of this type (e.g. \cite{Mat}), a
conservation law like (\ref{DivFree}) imposes differential
constraints on each moment. There are often algebraic couplings
between the various moments as well. This leads one to a system of
equations which are impossible to solve without assuming \textit{a
priori} that only some finite subset of the moments are important to
problem at hand. While adequate for many purposes, this removes any
possibility of reconstructing the current vector from its moments.
For example, if only a finite number of the moments were nonzero,
the associated vector field would involve the Dirac distribution and
its derivatives. Such current densities are clearly unphysical.


Dixon's moments avoid this type of problem by being chosen such that
there is exactly one differential equation imposed on them as a
consequence of (\ref{DivFree}). This equation is trivial, and only
states that the total charge does not change. The algebraic
constraints on the moments are also simple solve in closed form, so
it becomes possible to discuss the entire set of multipole moments.

To see how these simplifications work, first note that the moments
of some source would usually be defined by integrating it against a
suitable number of ``radial vectors'' referred to some arbitrarily
chosen origin. The moments should also be allowed to vary in time,
which is most naturally allowed by requiring that their defining
integrals be carried out only over spacelike hypersurfaces spanning
$W$. Before defining anything, we therefore need a reference frame
which foliates the worldtube with hypersurfaces $\Sigma(s)$. Each of
these must also be given an origin $z(s) \in \Sigma(s)$. Perhaps
imagining that these preferred points could represent the position
of a physical observer, we assume that they form a continuous
timelike worldline $Z \subset W$.

To be more specific, start by fixing the the reference worldline
$Z$. One may then specify the foliation by choosing a unit
future-pointing timelike vector field $n^{a}(s)$ on $Z$. $\Sigma(s)$
would be defined as the set of all points $x \in W$ such that
$n^{a}(s) \sigma_{a}(x,z(s)) = 0$ (see appendix \ref{Bitensors} for
the definition of $\sigma_{a}$). This naturally generalizes the
flat-spacetime concept of a hyperplane which is everywhere
orthogonal to $n^{a}$. In certain cases, these hypersurfaces will
intersect each other (inside $W$), and therefore fail to be a
foliation. Such instances would rarely occur in practice, however,
so we shall ignore them.

Completely specifying this reference frame also requires fixing the
``time'' parameter $s$ (up to an additive constant). We do this by
normalizing the tangent vector
\begin{equation}
v^{a}\big(z(s)\big) := \dot{z}(s) := \frac{\delta z^{a}(s)}{ds}
\label{vDefine}
\end{equation}
such that $v^{a} n_{a} = -1$.

One might be inclined to automatically choose $n^{a} = v^{a}$, but
more generality is needed for a number of problems in relativistic
mechanics. In particular, one might want to choose the reference
frame for the moments to be the center-of-mass frame of a body with
worldtube $W$ \cite{EhlRud,Dix79,CM}. In that case, $n^{a}$ would
point in the direction of the body's bulk momentum vector, while
$v^{a}$ would be the 4-velocity of its center-of-mass line. These
two objects are physically distinct, and would only coincide in
special cases.

Regardless, specifying $z(s)$ and $n^{a}(s)$ is sufficient to fix a
frame in which the multipole moments may be defined. All we need now
is some notion of a ``radial vector'' between $x \in \Sigma(s)$ and
$z(s)$. A particularly natural choice for this is the $X^{a}(x,z)$
defined in (\ref{XDefine}), which takes the form of a vector at $z$.
Accepting this, a multipole moment should then be defined by
integrating the current against a suitable number of these vectors.

But that cannot be done directly, as $J^{a'}(x) \notin T_{z}M$
unless $x = z$. It will be convenient to parameterize all possible
replacements for the current by a vector-valued distribution
$\hat{J}^{a}(X,z)$. We call this the current skeleton (although our
definition of it will differ slightly from Dixon's). Its precise
relation to $J^{a'}$ will be given later, but for now, we simply
assume that it vanishes outside of some finite neighborhood of
$X^{a}=0$. The $2n$-pole moment of the current is then defined to be
\begin{equation}
Q^{b_{1} \cdots b_{n} a}(z) := \int DX \, X^{b_{1}} \cdots X^{b_{n}}
\hat{J}^{a}(X,z) ~, \label{MomentDefine}
\end{equation}
where $DX := \sqrt{-g(z)} d^{4}X$. Note the moments are tensors on
$Z$. The skeleton is defined on the tangent bundle $TM$, and the
moments at $z$ are integrals over $T_{z}M$.

It might appear that the moments defined in this way depend on the
current at all times, but this is not true. The support of the
skeleton will be found to lie on the surface defined by $n^{a} X_{a}
=0$, which is simply $\Sigma(s)$. So one of the integrals in
(\ref{MomentDefine}) effectively drops out, and we fall back on the
expected three-dimensional definition of a multipole moment at some
chosen time.

Note that (\ref{MomentDefine}) trivially implies that for all $n
\geq 1$,
\begin{equation}
Q^{(b_{1} \cdots b_{n}) a} = Q^{b_{1} \cdots b_{n} a} ~.
\label{Constr1}
\end{equation}
This is the first of three index symmetries that must be satisfied
by the moments. The others are not so trivial, and form the main content of Dixon's formalism. We simply state that for all $n \geq 2$ \cite{Dix67, Dix70b, Dix74},
\begin{eqnarray}
n_{b_{1}} Q^{b_{1} \cdots b_{n-1} [b_{n} a]} &=& 0 ~,
\label{Constr2}
\\
Q^{(b_{1} \cdots b_{n})} &=& 0~.
\label{Constr3}
\end{eqnarray}
The monopole moment also has the special form
\begin{equation}
Q^{a}(z) := q v^{a}(z) ~,
\label{Monopole}
\end{equation}
where $q$ is the total charge defined by (\ref{chargeDefine}).

We now need to relate $\hat{J}^{a}$ to $J^{a'}$. In order to do so,
it will be convenient to think of the current skeleton as a linear
functional on the space of all $C^{\infty}$ test functions with
compact support (as is typical in distribution theory). Given some
$\Phi^{a}(X,z)$ related to a $\phi^{a'}(x)$ in this class, we want
to know the behavior of
\begin{equation}
\int DX \, \hat{J}^{a}(X,z) \Phi_{a}(X,z)
\label{SkeletonFunctional}
\end{equation}
for all possible forms of $\phi^{a'}$. This functional should be
related in some simple way to
\begin{equation}
\left\langle \mathbf{J}, \bm{\phi} \right\rangle := \int d^{4}x
\sqrt{-g(x)} \, J^{a'} \phi_{a'} ~,
\end{equation}
which determines $J^{a'}$. Note that the (\ref{SkeletonFunctional})
depends on $z$, while $\left\langle \mathbf{J}, \bm{\phi}
\right\rangle$ does not. It is therefore simplest to link the two
functionals together by simply integrating out the $z$-dependence:
\begin{equation}
\left\langle \mathbf{J}, \bm{\phi} \right\rangle = \int ds \! \int
DX \, \hat{J}^{a}(X,z) \Phi_{a}(X,z)~. \label{JtoJhat}
\end{equation}

Explicitly relating $\phi^{a'}(x)$ and $\Phi^{a}(X,z)$ would now
complete the definition of the skeleton in terms of $J^{a'}$. It is
natural to suppose that $x=\exp_{z} X$, but there must also be a
propagator which maps vectors at $z$ into vectors at $x$. The
convenient choice for this happens to be the bitensor $H^{a'}{}_{a}$
defined by (\ref{HDefine}). This is simply the identity if the
$X^{a}$ are used as coordinates (with a fixed $z$). In general,
though,
\begin{equation}
\Phi_{a}(X,z) := H^{a'}{}_{a}(\exp_{z} X, z) \phi_{a'}(\exp_{z} X)
~. \label{phiPhi}
\end{equation}

Taken together, the preceeding equations completely define the
moments (or equivalently the skeleton) of the current vector. To see
that they automatically imply (\ref{DivFree}), choose a scalar test
function $\varphi$, and then set $\phi_{a'} = \nabla_{a'} \varphi$.
With this definition,
\begin{equation}
\Phi_{a}(X,z) = \frac{\partial}{\partial X^{a}} \varphi(\exp_{z} X)
~.
\label{LittlePhi}
\end{equation}

For any object depending on both $X^{a}$ and $z$, operator
$\partial/ \partial X^{a}$ actually has a very natural
interpretation in the theory of vector bundles. It is called a
vertical covariant derivative in \cite{Dix74, Dix79}, and was given
the special symbol $\nabla_{*a}$. Using this notation,
(\ref{LittlePhi}) may be rewritten as
\begin{equation}
\Phi_{a} = \nabla_{*a} \varphi ~.
\end{equation}
Combining this with (\ref{JtoJhat}),
\begin{equation}
\left\langle \nabla \cdot \mathbf{J}, \varphi \right\rangle = \int
ds \int DX \left( \nabla_{*a} \hat{J}^{a} \right) \varphi ~.
\label{DivFreeProve1}
\end{equation}

The divergence on the right-hand of this equation may be evaluated
by defining the Fourier transform of the current skeleton. Let
\begin{equation}
\mathcal{F}[\hat{\mathbf{J}}]^{a}(k,z) := \int DX \hat{J}^{a}(X,z)
e^{-i k_{b} X^{b}} ~,
\end{equation}
where $k^{a} \in T_{z}M$. It then follows immediately from
(\ref{MomentDefine}) that
\begin{equation}
\mathcal{F}[\hat{\mathbf{J}}]^{a}(k,z) = \sum_{n=0}^{\infty}
\frac{(-i)^{n}}{n!} k_{b_{1}} \cdots k_{b_{n}} Q^{b_{1} \cdots b_{n}
a} (z) ~.
\end{equation}
So the Fourier transform of the current skeleton is a generating function for the moments.

It will be convenient to define a reduced current skeleton
$\check{J}^{a}(X,z)$ that drops the first (monopole) term in this
series: $\mathcal{F}[\check{\mathbf{J}}]^{a} =
\mathcal{F}[\hat{\mathbf{J}}]^{a} - Q^{a}$. Noting that
$\mathcal{F}[\nabla_{*a} A^{a}]=i k_{a} \mathcal{F}[\mathbf{A}]^{a}$
for any $A^{a}(X,z)$, (\ref{Constr3})  immediately implies that
\begin{equation}
\nabla_{*a} \check{J}^{a} = 0 ~. \label{ReducedSkel}
\end{equation}
The divergence of the full current skeleton therefore depends only
on the monopole moment. Using the identity
$\mathcal{F}[\delta(X)]=1$, we see that
\begin{equation}
\nabla_{*a} \hat{J}^{a} = Q^{a} \nabla_{*a} \delta(X) ~.
\end{equation}
Substituting this result into (\ref{DivFreeProve1}), and using
(\ref{Monopole}) finally shows that $\left\langle \nabla_{a'}
J^{a'}, \varphi \right\rangle =0$ for any $\varphi$. Hence,
(\ref{DivFree}) is always satisfied, as claimed.

\end{document}